*Daniel Treisman*

**ELECTIONS IN RUSSIA, 1991-2008**

University of California, Los Angeles

Moscow

State University - Higher School of Economics

2009

**Treisman D. Elections in Russia, 1991 - 2008:** Working paper WP7/2009/06. – Moscow: State University — Higher School of Economics. - 38 p.

      In this paper, I review the main trends in voting in national elections in Russia since 1991, discuss the evidence of manipulation or falsification by the authorities, and use statistical techniques to examine the determinants of voting trends. The emphasis is on explaining change over time in the vote shares of different parties or groups of parties, not on identifiying social, economic, or opinion correlates of individuals' party choice at a given point in time—a subject that has been well-studied in previous work by various scholars. My goal is to provide a relatively comprehensive introduction to the changing electoral landscape in the two decades since the birth of the independent Russian state. The paper offers a more detailed and technical examination of the evidence that is summarized in Chapter 10 of my book manuscript titled *The Return: Russia's Journey from Gorbachev to Medvedev*.

*Treisman Daniel - University of California, Los Angeles*

# The patterns

Between 1989 and 2008, Russians had the chance to vote in seven elections to the national parliament, five presidential elections, four referenda, and several rounds of voting for regional legislatures and governors.[1]

All the national elections pitted multiple candidates against each other. From 1993, many parties competed. In elections to the lower house of parliament, the Duma, between 1993 and 2003, each voter cast two ballots.[2] One was for a candidate to represent the local district; such candidates made up half of the Duma's members. The second was for a national party; each party that won at least five percent of the votes received seats proportional to its vote share. This party-list voting determined the second half of the Duma's members. From 2007, the single-member districts were abolished and voters only voted on the party-list ballot. At the same time, the threshold for representation was raised to seven percent of the votes. In presidential elections, if no candidate received 50 percent of the votes, a second round was held several weeks later between the two top vote-winners from the first round.

Since 1993, parties and electoral blocs have appeared, disappeared, merged, split, and renamed themselves many times. Despite this, one can sort the parties into rough families based on what policies they support, and examine how different groups of parties have fared at the polls. The policy positions of the main parties have differed on two main dimensions, and sometimes these collapse to just one.[3] First, parties have differed in their attitudes towards market-oriented economic reform. Second—and this dimension sometimes lines up with the first—parties have adopted different positions on the tradeoff between individual rights and the authority of the state.

At the anti-reform end of the spectrum were a number of communist blocs, led from 1993 by the Communist Party of the Russian Federation (CPRF). In close accord with the CPRF was the smaller Agrarian Party, which represented the old collective farm directors. Further to the left were various small revolutionary Marxist groups. Somewhat less opposed to economic reform, but more hostile than the Communists towards individual rights, was the perversely misnamed Liberal Democratic Party of Vladimir Zhirinovsky. It favored an authoritarian, imperialistic state that would provide many of the social benefits promised under Communism. Together, these two groups constituted the extreme opposition to Yeltsin and his pro-reform governments.

At the other end of the spectrum, a number of parties supported both economic reforms and the protection of individual rights. The most economically liberal of these was a bloc led by Yegor Gaidar, Anatoly Chubais, and later Boris Nemtsov, named successively "Russia's Choice," "Russia's Democratic Choice," and the "Union of Right Forces." A second party that favored economic reforms and individual rights was named Yabloko. Led by an ambitious, pro-market economist, Grigory Yavlinsky, its positions were similar to those of the Gaidar liberals, except that it criticized Yeltsin's governments rather than supporting them and was more inclined to favor socially oriented policies even at the cost of high inflation. Between these two extremes lay a fluid band of "center" parties, moderate nationalist groups, special interest blocs, and organizations whose main purpose was to support officials currently in office. The latter, known as "parties of power," included the "Our Home is Russia" (OHIR) bloc set up by Prime Minister Viktor Chernomyrdin in 1995 and the Unity party (later renamed United Russia) created for Putin in 1999.

---

[1] Elections for governors were eliminated at the end of 2004.

[2] In fact, in 1993, there were four ballots: two for the Duma; one for the Council of Federation, which was no longer elected after 1993; and one on the referendum on approving the draft constitution.

[3] This is discovered, for instance, by analyzing how the members of the different parties have voted in the parliaments since 1994. The patterns of voting can be well summarized in just two dimensions. See F. T. Aleskerov, N. Yu. Blagoveshchensky, G. A. Satarov, A. V. Sokolova, V. I. Yakuba, *Vliyanie i strukturnaya ustoichivost v Rossiiskom parlamente (1905-1917 i 11993-2005 gg.)*, Moscow: Fizmatlit, 2007, especially at p.113, where the authors graph the early positions of the main parties. These authors interpret the two dimensions as representing loyalty vs. opposition to the regime and ideology vs. pragmatism. However, attitudes towards economic reform and on individual rights vs. state authority seems to me to fit better.

Figures 1-3 show how these families of parties have performed in successive national elections. In each graph, the dashed line indicates the share of the valid vote won by parties from the relevant group in the party-list part of the Duma elections; the solid line records the percentage won by candidates from the relevant group of parties in presidential elections.[4]

It does not take long to see the pattern. Support for the liberal reformers fell sharply during the 1990s. (Since President Yeltsin appealed to both pro-reform and centrist voters, one can see this most accurately focusing on the parliamentary votes.) In Duma elections, the vote for liberal reform parties fell from 34 percent of the valid vote in 1993 to less than four percent in 2007. That represented a drop from 18 million to fewer than three million pro-reform voters. Support for the extreme opposition—Communists and the LDP—peaked in the 1995 Duma election at 44 percent (30 million voters), and then fell gradually, reaching 22 percent (15 million voters) in 2007.[5] (The extra dip in 2004 most likely reflects the fact that both Zhirinovsky and the Communist leader, Gennady Zyuganov, chose not to run for president that year and instead nominated little-known, uncharismatic colleagues.) The great beneficiaries of the shrinking pro-reform and extreme opposition votes were the party of power and the Kremlin-supported presidential candidates, whose support fell at first, but then rallied from 1999, reaching about 70 percent of the vote—more than 50 million voters—in 2008.

---

[4] In 1996, when there were two rounds, I show the result for the first round.

[5] There were changes in the composition of the extreme opposition vote: in 1993, Zhirinovsky's LDP surged ahead of the Communists. This reversed in the 1995 election.

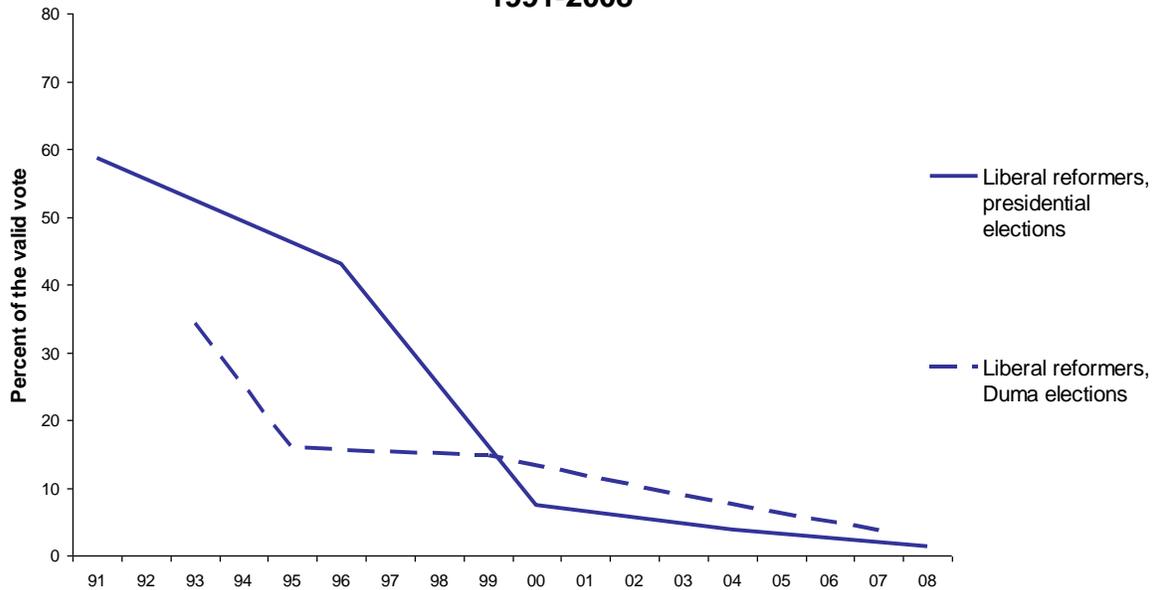

**Figure 1.   Voting for liberal reformers in Russian elections, 1991-2008**

Source: Central Electoral Commission. "Liberal reformers" are: 1991: Yeltsin; 1993: Russia's Choice, Yabloko, PRES, RDDR; 1995: Yabloko, Russia's Democratic Choice, Forward Russia, Pamfilova-Gurov-Lysenko Bloc, Common Cause, PRES, Party of Economic Freedom; 1996: Yeltsin, Yavlinsky; 1999: Union of Right Forces, Yavlinsky; 2000: Yavlinksy, Titov, 2003: Yabloko, SPS, Razvitie Predprinimatelstva; 2004: Khakamada; 2007: Grazhdanskaya Sila, Union of Right Forces, Yabloko; 2008: Bogdanov. 1996 election is first round.

**Figure 2.   Voting for the extreme opposition in Russian elections, 1991-2008**

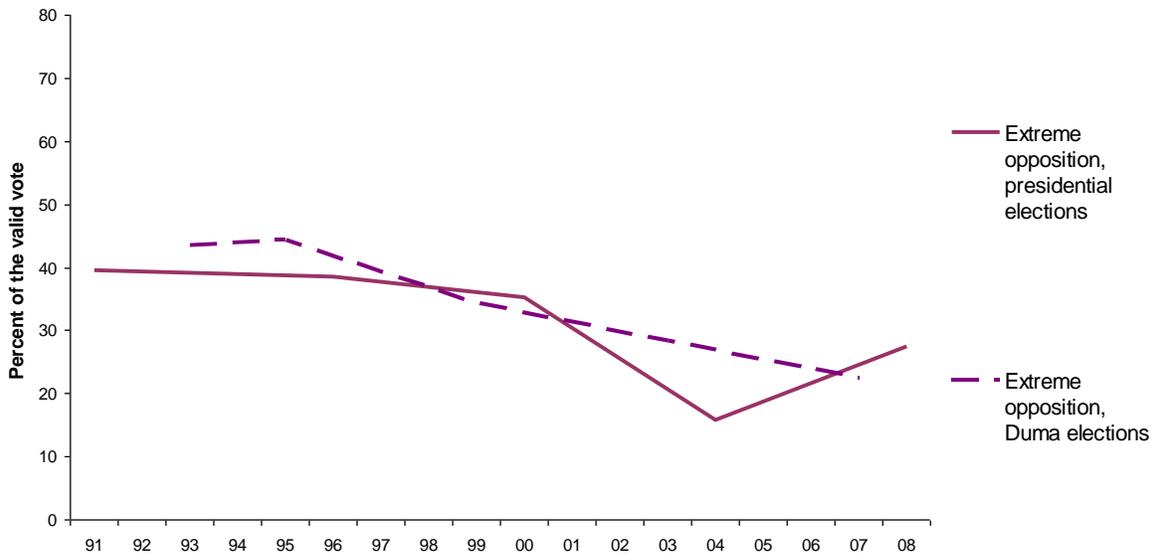

Source: Central Electoral Commission. "Extreme opposition" is "Communists" + "LDP". "Communists" are: 1991: Ryzhkov, Tuleev, Makashov, Bakatin; 1993: KPRF and Agrarians; 1995: KPRF, Agrarians, Power to the People, Communists-Working Russia; 1996: Zyuganov; 1999: KPRF, "Stalinist Bloc--For the USSR," "Communists, Working Russia," Socialist Party of Russia, Russian Socialist Party; 2000: Zyuganov, Tuleev; 2003: KPRF, Agrarians; 2004: Kharitonov; 2007: KPRF, Agrarians; 2008: Zyuganov. "LDP" is Liberal Democratic Party of Russia, Zhirinovsky, or Malyshkin (2004). 1996 election is first round.

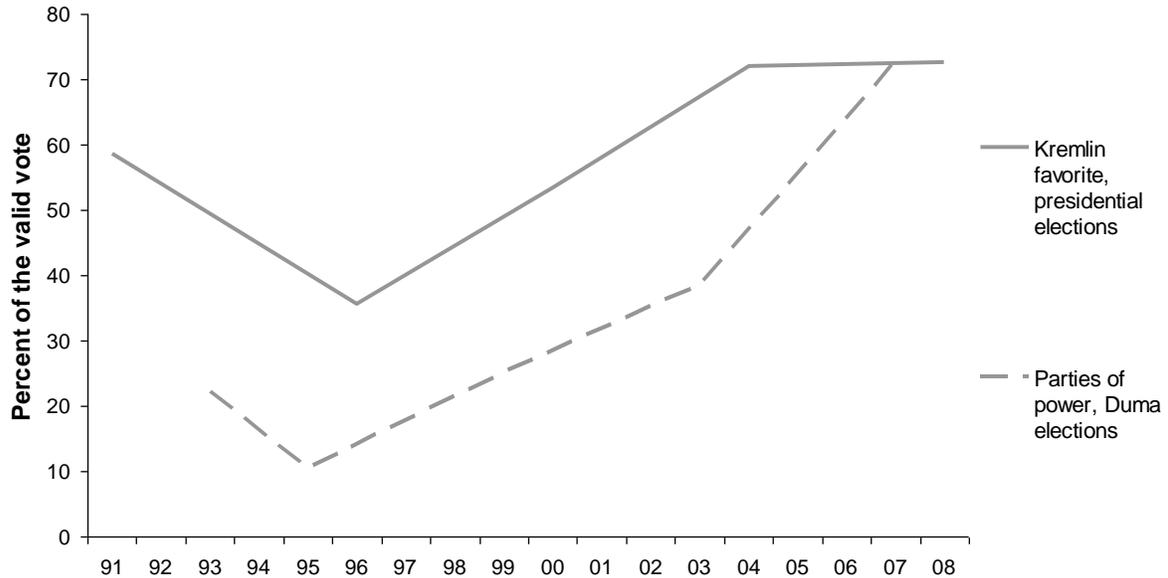

**Figure 3. Voting for "Party of Power" or Kremlin favorite in Russian elections, 1991-2008**

Kremlin favorite, presidential elections

Parties of power, Duma elections

Source: Central Electoral Commission. "Kremlin favorites" are: 1991: Yeltsin; 1996: Yeltsin; 2000: Putin; 2004: Putin; 2008: Medvedev. "Parties of power" are: 1993: Russia's Choice, PRES; 1995: Our Home is Russia (OHIR); 1999: Unity, OHIR; 2003: United Russia; 2007: United Russia, Just Russia. 1996 election is first round. By "Kremlin favorite," I mean favorite of the incumbent *Russian* authorities (and not the Soviet authorities in 1991).

Consider also the geography of support for candidates from different ideological groupings. Figures 4-6 show the patterns of regional voting for the incumbent and for the Communist opposition in the 1991, 1996 (first round) and 2004 presidential elections. In 1991, there is no obvious pattern in either Yeltsin's or his Communist rivals' votes. The strongholds of each are scattered across the map in apparently random clumps. By 1996, this has changed markedly: a North-South divide has emerged. Yeltsin's support is strongest in the North, and his Communist challenger pulls in relatively more votes in a belt of regions along the country's extended underbelly, from Smolensk in the West to Amur in the Far East. A similar North-South divide could be seen in the April 1993 referendum on confidence in Yeltsin, and in the votes for reformers and Communists in the 1993 and 1995 parliamentary elections. Russia's regions differ in latitude by almost 27 degrees, from Dagestan, which is on a level with Southern France, to the Taimir Autonomous Okrug, situated further north than Fairbanks, Alaska. In 1996, the vote for Yeltsin was about 0.84 percentage points higher for every degree further north the region was located.

However, by 2004 the pattern had changed again. The Communist challenger still did relatively better in the so-called "red belt" of the South and South-West. But unlike Yeltsin's, Putin's regional support was not concentrated in the North. Although it is not obvious from just looking at Figure 6, there is a powerful underlying logic. Putin's strongest showings were all in the ethnically non-Russian areas of the country.[6] The 18 regions with the highest votes for Putin in 2004 were all autonomous republics or autonomous districts, named after some non-Russian nationality (I will call such regions "ethnic regions"). Seven of these reported votes for Putin of more than 90 percent. Among the *non*-ethnic regions, there remained a North-South gradient in pro-Putin voting that was almost as strong as under Yeltsin. For every degree further north a non-ethnic region was located, the vote for Putin was .65 percentage points higher, compared to .84 points under Yeltsin. Among ethnic republics, by contrast, the vote for Putin increased as one went further south.

---

[6] See Christopher Marsh and James Warhola, "Ethnicity, Ethno-territoriality, and the Political Geography of Putin's Electoral Support," *Post-Soviet Geography and Economics*, 2001, 42, 4, pp. 1-14.

Over time, voters in the ethnic regions appear to have become ever more enamored of the Kremlin's candidates. In 1991, the ethnic regions had actually voted strongly for the Communists; Yeltsin—the serving chairman of the Russian parliament, and so Russia's chief executive—polled seven percentage points lower there than elsewhere. But by the first round of the 1996 election, this had reversed: Yeltsin's vote was now almost eight percentage points higher in the ethnic regions. In 2000, 2004, and 2008, the Kremlin-favored candidate won 7, 12, and 7 percentage points more, respectively, in the ethnic regions. The pro-incumbent advantage was even stronger for Putin's United Russia party, which polled more than 14 percentage points higher on average in the ethnic regions in both 2003 and 2007.[7]

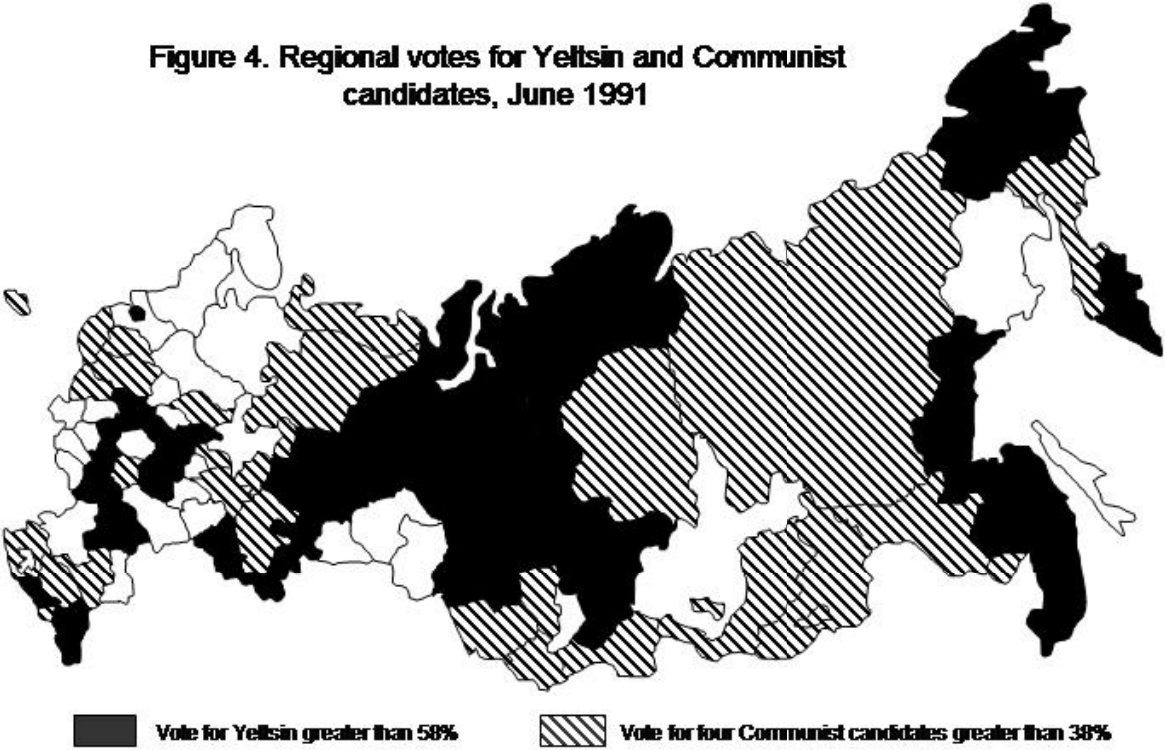

**Figure 4. Regional votes for Yeltsin and Communist candidates, June 1991**

Vote for Yeltsin greater than 58%          Vote for four Communist candidates greater than 38%

Source: Author's calculations from Michael McFaul and Nikolay Petrov, *Politichesky almanakh Rossii 1995*, Moscow: Moscow Carnegie Center, 1995, pp.655-6. Percent of the valid vote.

---

[7] My calculations from Central Electoral Commission data; these figures refer to the difference in the average result between ethnic and non-ethnic regions.

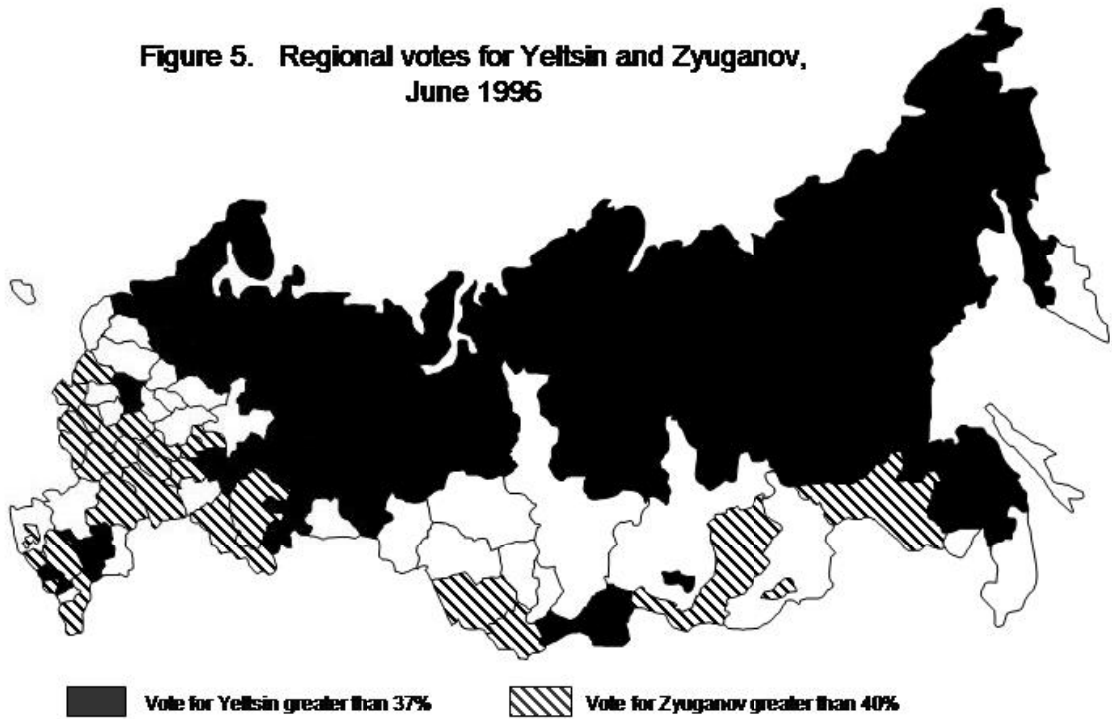

**Figure 5. Regional votes for Yeltsin and Zyuganov, June 1996**

■ Vote for Yeltsin greater than 37%    ▨ Vote for Zyuganov greater than 40%

Source: Author's calculations from data downloaded from Central Electoral Commission of the Russian Federation. Percent of the valid vote.

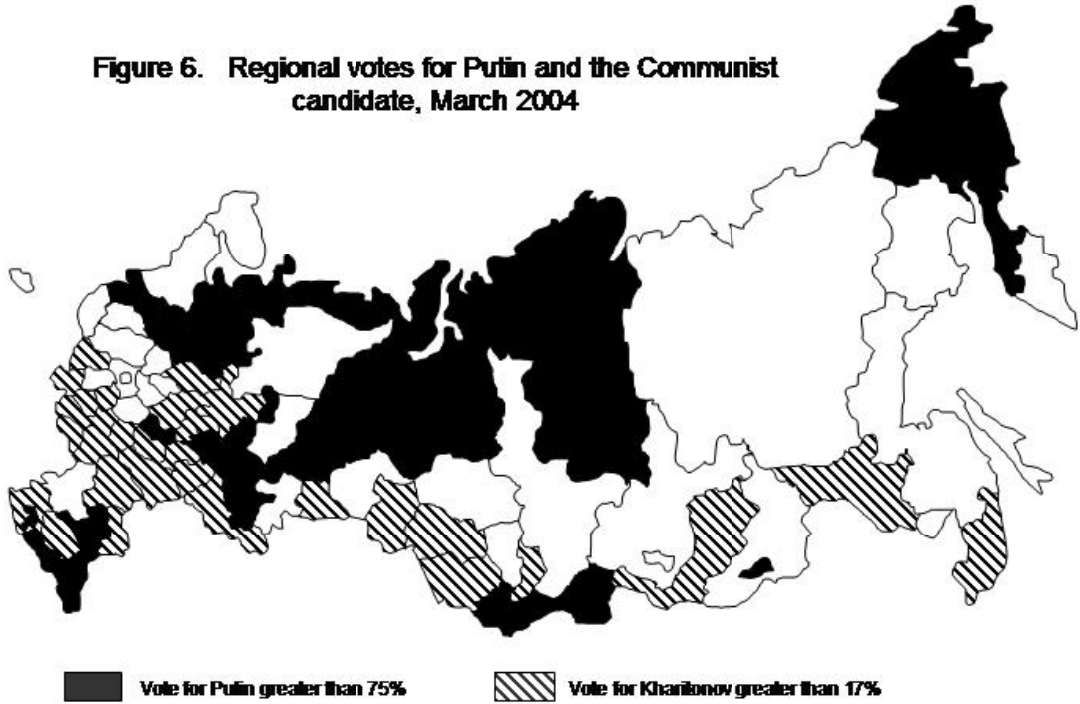

**Figure 6. Regional votes for Putin and the Communist candidate, March 2004**

■ Vote for Putin greater than 75%    ▨ Vote for Kharitonov greater than 17%

Source: Author's calculations from data downloaded from Central Electoral Commission of the Russian Federation. Percent of the valid vote.

Thus, we have several puzzles to explain. First, why did electoral support for the liberal reformers decline from the early 1990s? Second, why did support for the extreme opposition at first rise and then fall from the late 1990s? Third, why did Kremlin-favored parties and candidates do worse in the early 1990s, but then surge ahead after 1999? Fourth, why did a North-South divide emerge in the mid-1990s between support for reformist incumbents like Yeltsin and Communists like Zyuganov? Finally, why did Putin and the pro-Kremlin parties do so well after 2000 in the ethnic regions?

Manipulation and Fraud

A first question is whether the trends observable in support for the various parties simply reflect the effects of increasing manipulation and fraud by state authorities.

I will not review here the copious reports of electoral abuses in Russia, especially since 1999. These have been discussed in great detail in all the major newspapers. Scholars have also detailed the various anomalies to be found in the official statistics on voting. Such anomalies do not prove that abuses occurred; innocent explanations are conceivable. But, in association with the journalists' accounts, they paint a disturbing picture.

Most of these oddities have to do with the reported levels of voter turnout. First, a growing number of Russia's 89 regions—and of the roughly 2,700 rayons (or districts) within them—have been reporting extremely high turnout (see Table 1, rows 1 and 2). By 2008, 36 towns and rural districts—mostly in the republics of Bashkortostan, Tatarstan, and Tyva—had turnout higher than 99 percent. Second, the vote for the Kremlin's favored candidate or party has, since 2000, been much higher in these regions and rayons.

Table 1.   Turnout and voting for different parties in Russia's regions and rayons

| Election | 1991 | 1993 | 1995 | 1996 I | 1996 II | 1999 | 2000 | 2003 | 2004 | 2007 | 2008 |
|---|---|---|---|---|---|---|---|---|---|---|---|
| 1. Number of regions where turnout was greater than 85 percent | 3 | 0 | 0 | 0 | 0 | 0 | 2 | 1 | 8 | 9 | 10 |
| 2. Number of rayons (counties) where turnout was greater than 85 percent | | | 115 | 142 | 155 | 114 | 191 | 172 | 408 | | 616 |
| 3. Correlation among regions between turnout and vote for | | | | | | | | | | | |
| Kremlin candidate/party [a] | -.11 | -.53 | -.13 | -.28 | -.12 | -.01 | **.48** | **.68** | **.69** | **.90** | **.80** |
| KPRF [b] | 0.12 | **.41** | **.37** | **.41** | .16 | .02 | -.22 | -.29 | -.46 | -.73 | -.72 |
| Fatherland-All Russia | | | | | | **.41** | | | | | |
| 4. Correlation among rayons (counties) between turnout and vote for | | | | | | | | | | | |
| Kremlin candidate/party | | | .07 | -.23 | -.13 | -.08 | **.38** | **.63** | **.56** | n.a. | **.71** |
| KPRF | | | **.29** | **.41** | .15 | .08 | -.09 | -.19 | -.24 | n.a. | -.60 |
| Fatherland-All Russia | | | | | | **.43** | | | | | |

Sources: data downloads from Central Electoral Commission of the Russian Federation, Michael McFaul and Nikolay Petrov, *Politichesky almanakh Rossii 1995*, Moscow: Moscow Carnegie Center, 1995; Robert Orttung and Scott Parish, "Duma Votes Reflect North-South Divide," *Transition*, 23 February, 1996, pp.12-14. [a] in 1991, Yeltsin; [b] in 1991, Ryzhkov, Bakatin, Tuleev, and Makashov. Positive correlations greater than 0.2 in bold.

Since 2000, higher turnout has correlated positively—and in later years, very strongly—with higher votes for the Kremlin's favorites (Table 1, rows 3 and 4).

Not only has very high turnout become more frequent and correlated with pro-Kremlin voting, the distribution of turnout across regions and districts has become quite peculiar from a statistical point of view.[8] In most elections where districts are relatively similar, the distribution of turnout across the districts approximates the Normal Distribution, with most districts bunching around the average level of turnout and far fewer having either relatively high or relatively low turnout. An almost perfect example of this is the distribution of turnout in different districts of Moscow in the 1995 parliamentary election (Figure 7A).

But compare this to Figure 7B, which graphs the distribution of turnout for urban districts in the ethnic republics in 2004. The bell has been transformed into a camel's back with two ungainly humps. That year, there were two main groups of urban districts in the ethnic republics—one group where turnout averaged 60-65 percent, close to the national average, and another where turnout averaged 90 percent. There could be some innocent reason why some of these districts had much more enthusiastic voters than others. But it is not obvious what that reason might be. Another possibility is that in some—but not all—of the urban districts in the ethnic republics, voters were pressured to vote or ballots were stuffed on behalf of non-voters, resulting in extremely high turnout.

**Figure 7A.   Turnout in Moscow's 121 districts, 1995**

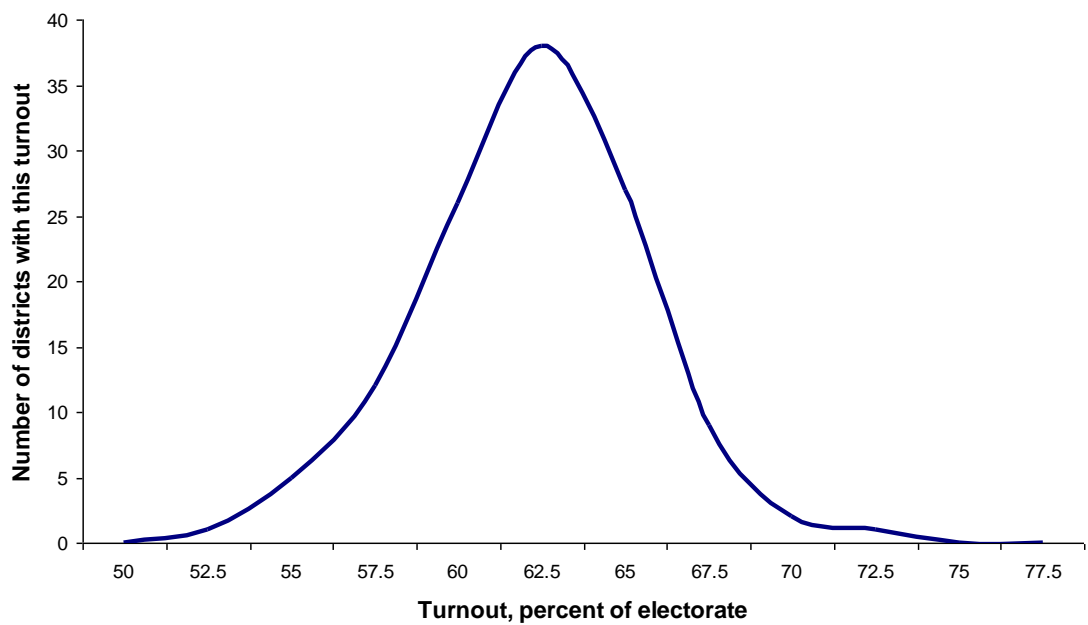

Source: Calculated from data of the Central Electoral Commission of the Russian Federation. Smoothed line.

---

[8] For detailed analyses of this, see the work of Mikhail Myagkov, Peter Ordeshook and their collaborators, in particular: Mikhail Myagkov, Peter Ordeshook and Dmitrii Shakin, "Fraud or Fairytales? Russian and Ukrainian Electoral Experience", *Post-Soviet Affairs*, 21, 2, April-June 2005, pp. 91-131, and Andrei Beryozkin, Mikhail Myagkov and Peter Ordeshook "Location and Political Influence: A Further Elaboration of Their Effects on Voting in Recent Russian Elections." *Eurasian Geography and Economics*, 44, 3, March 2003, pp. 169-183.

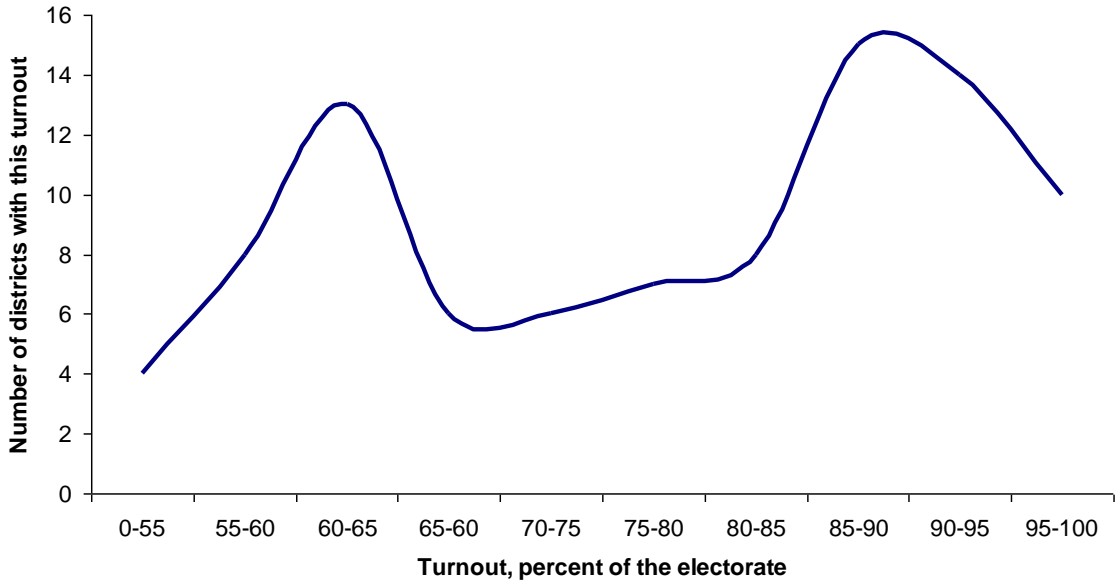

**Figure 7B.   Turnout in urban districts of ethnic republics, 2004**

Source:  Calculated from data of the Central Electoral Commission of the Russian Federation. "Urban districts" are those where 100 percent of the population is urban. Smoothed line.

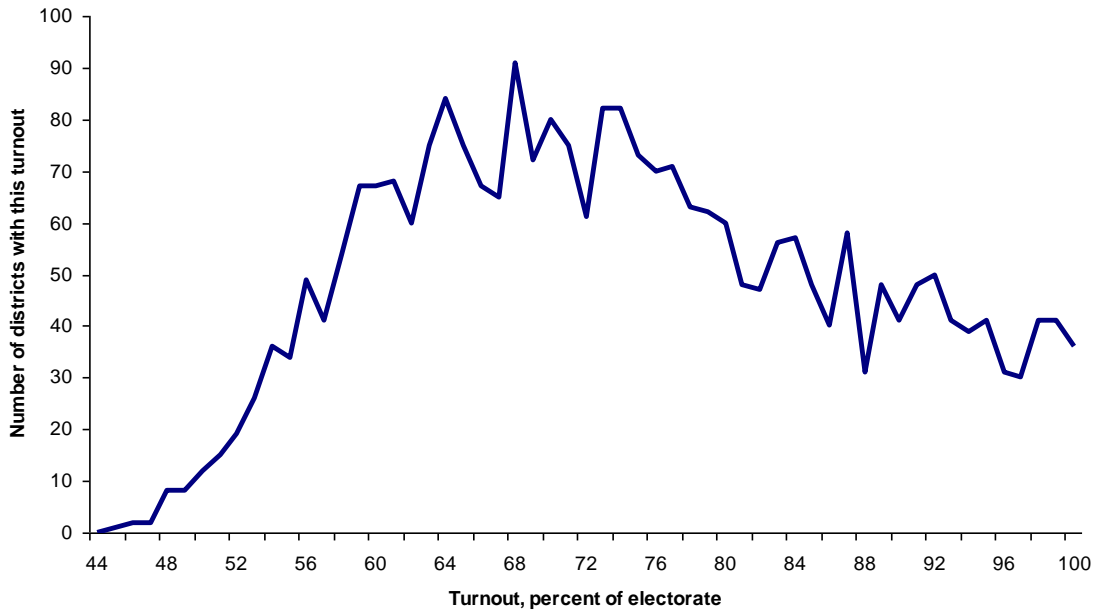

**Figure 7C.   Turnout in all 2,749 districts, 2008 election**

Source: Calculated from data of the Central Electoral Commission of the Russian Federation.

If we look at turnout across all rayons for the 2008 election, in place of a bell-shaped curve, we see a mountain with numerous jagged indentations, and a right tail that is much higher than the left tail (Figure 7C). As various statistically-minded observers have noted, such jagged indentations could plausibly be caused by rounding up the level of turnout to nice round numbers. If one plots the distribution across all 94,000 polling stations (not shown here), it turns out that there are abnormal spikes in the numbers of precincts where turnout was an exact multiple of five.[9]

In short, voter turnout has been abnormally high in a growing number of regions and rayons. Since 2000, such high turnout has been associated with higher votes for the parties and candidates favored by the central authorities. It has also tended to occur in regions where there were numerous allegations of electoral fraud—in particular, the ethnic regions. Although innocent explanations are possible, the most plausible one is that officials have pressured local voters to turn out in high numbers and vote for central incumbents, or stuffed the ballot boxes, or rewritten the electoral protocols. Under this interpretation, extremely high turnout becomes an indicator that irregularities have occurred.

If one accepts this interpretation, one can make inferences about which parties benefited from fraud or pressure in particular elections. The beneficiaries would be those parties whose vote shares were higher where turnout was higher. After 1999, that was the Kremlin's allies. Between 2000 and 2008, a high turnout correlated with a high vote for Putin, Medvedev, or United Russia (see Table 1, rows 3 and 4). In 2007, the correlation reached 0.90; for each additional ten percentage points of turnout in a region, the United Russia vote was on average nine percentage points higher.

The 1990s were quite different. In 1993, 1995, and the first round of the 1996 presidential election, higher turnout correlated with a higher vote share for the *Communists*, and a lower vote for Yeltsin or the pro-Kremlin parties. (This relationship became much weaker in the second round of 1996, as votes in some high turnout regions like Tatarstan shifted from Zyuganov to Yeltsin.) An obvious hypothesis is that regional and local officials were using the familiar administrative tricks on behalf of the Communists. This was exactly what Yeltsin's campaign workers believed at the time. Sergey Lissovsky, who was organizing tours of rock musicians for Yeltsin, said that many of the regional leaders, expecting Zyuganov to win, were deliberately sabotaging Yeltsin's campaign. "In the provinces, they arrested our vehicles containing equipment and printed materials and cut off electricity to the squares." In Lissovsky's estimate: "Half the 'resources' were initially against Yeltsin. Of the remaining half, one third were neutral, not supporting either side."[10]

Then in 1999, Moscow's mayor Yury Luzhkov, together with the presidents of Tatarstan and Bashkortostan, forged a powerful coalition of regional governors to contest the Duma elections. Their "Fatherland-All Russia" bloc was thought to be the main rival to the Communists and Putin's Unity. And it was Fatherland-All Russia that received higher votes that election in the regions and rayons where turnout was higher (Table 1). The regional bosses were apparently using their administrative resources on Fatherland's behalf. That did not save it from trailing Unity in the national totals, and it surrendered later, merging with Unity to form United Russia in 2001.

---

[9] Nabi Abdullaev, "Medvedev won by curious numbers," *The Moscow Times*, April 14, 2008, p.1. Another type of statistical oddity occurred in some ethnic regions in 1996, where in some districts thousands of voters apparently voted for Zyuganov in the first round of the presidential election but switched to voting for Yeltsin *against* Zyuganov in the second. See, for instance, Valentin Mikhailov, "Regional elections and democratisation in Russia," in Cameron Ross, ed., *Russian Politics under Putin*, New York: Manchester University Press, 2004, pp.198-220.

[10] Ekaterina Deeva, "Russkaya ruletka—96. Kak delali prezidenta," *Moskovskiy Komsomolets*, 6 July, 2001, interviews with Korzhakov, Sergey Lissovsky, and Sergey Zverev. Beryozkin, Myagkov and Ordeshook ("Location and Political Influence") also conclude that the voting data make the most sense if one assumes that fraud in 1996 redounded to the Communists' benefit.

All the indications are that manipulation and fraud increased over time, becoming quite widespread under Putin. Can this explain the pattern of results? Was it electoral foul play that caused the decline in the vote shares of the reformers and the extreme opposition, while boosting the share of Kremlin-connected parties and candidates? Based on the various facts presented so far, one would expect the answer to be yes.

Here things get more puzzling. If the official results of elections reflect major falsification and pressures on voters, one should expect these results to diverge noticeably from those of public opinion polls. And yet a variety of polling organizations, some with strong reputations for independence, have been able to predict the election results accurately on the basis of their polls. Table 2 shows the forecasts of a few of the main pollsters on the eve of each election, along with the actual election results.

Table 2.   Forecasts of pollsters compared to election results

|  | 1993 | 1995 | 1996, I | 1999 | 2000 | 2003 | 2004 | 2007 | 2008 |
|---|---|---|---|---|---|---|---|---|---|
| **Percent of vote for Kremlin favorite** | *Russia's Choice* | *OHIR* | *Yeltsin* | *Unity* | *Putin* | *United Russia* | *Putin* | *United Russia* | *Med-vedev* |
| *Forecasts of:* | | | | | | | | | |
| FOM | | 9.2 | 34 | 20 | 53 | 33.6 | 73 | 63 | 70.0 |
| VCIOM | 15 | 10 | 35-9 | 17 | 53-55 | 33-35 | 73.6 | 62.1 | 72.6 |
| Levada Center | | | | | | 33.7 | 73.7 | 62.8 | 73.8 |
| Average | 15 | 9.6 | 35.5 | 18.5 | 53.5 | 33.8 | 73.4 | 62.6 | 72.1 |
| *Actual percent of total votes* | *14.5* | *10.1* | *35.1* | *23.3* | *52.9* | *37.6* | *71.3* | *64.3* | *70.3* |
| **Turnout** | | | | | | | | | |
| *Forecasts of:* | | | | | | | | | |
| FOM | | 64 | 70 | 65 | 67 | 64-5 | 60 | 58-9 | 68.7 |
| VCIOM | | | | 60 | 59 | 56 | 60-65 | 53.6 | 69.7 |
| Levada Center | | | | 64-5 | | 60 | 62 | 53.4 | 62.3 |
| Average | | 64 | 70 | 62.5 | 63.5 | 60.2 | 61.5 | 55.2 | 66.9 |
| *Actual percent of eligible voters* | *54.3* | *64.4* | *69.7* | *61.7* | *68.9* | *55.7* | *64.3* | *63.7* | *69.7* |

Sources: **1993:** VCIOM: Yury Levada, *Nashi dyesat let: itogi i problyemy*, http://www.levada.ru/levadaocherki.html, p.248.
**1995:** FOM: http://bd.english.fom.ru/report/map/oslon/ed034726. VCIOM: Yury Levada, *Nashi dyesat let: itogi i problyemy*, http://www.levada.ru/levadaocherki.html, p.249. **1996:** FOM: http://bd.english.fom.ru/report/map/oslon/ed034726. VCIOM: Yury Levada, *Nashi dyesat let: itogi i problyemy*, http://www.levada.ru/levadaocherki.html, p.249. **1999:** FOM: http://bd.english.fom.ru/report/map/oslon/ed034726.

VCIOM: http://www.yabloko.ru/Press/1999/9912154.html. **2000:** FOM: http://bd.english.fom.ru/report/map/oslon/ed034726. VCIOM: http://www.polit.ru/news/2000/03/22/ (21 March 2009, Polit.ru). **2003:** Levada: party vote is percent of those intending to vote, from Nov 13-16 survey, adjusted to exclude those planning to vote but who say "don't know" for whom, http://www.levada.ru/press/2003112104.html. FOM: http://bd.english.fom.ru/report/map/oslon/ed034726. VCIOM: http://wciom.ru/arkhiv/tematicheskii-arkhiv/item/single/29.html?no_cache=1&cHash=febd77c4a5. **2004:** Levada: http://www.levada.ru/press/2004031802.html.
FOM: http://bd.fom.ru/report/map/o040802. VCIOM: http://wciom.ru/arkhiv/tematicheskii-arkhiv/item/single/623.html?no_cache=1&cHash=3944901775
**2007:** Levada: http://www.levada.ru/press/2007120301.html. FOM: http://bd.fom.ru/report/cat/prognoz261107. VCIOM: http://wciom.ru/arkhiv/tematicheskii-arkhiv/item/single/9240.html?no_cache=1&cHash=378e108cc3. **2008:** All: http://www.cikrf.ru/rcoit/news/politprogn_2008.pdf. FOM on turnout: http://bd.fom.ru/report/cat/elect/pres_el/president_elect_2008/prognozitog08
**Electoral results:** Central Electoral Commission of the Russian Federation.

For the most part, the forecasts of the performance of Kremlin-connected contestants have been on target. Unity and UR do exceed the average forecast by up to five points in 1999, 2003, and 2007. But the forecasts are somewhat too high for Putin and Medvedev in the presidential contests.[11] Overall, had votes in elections for the Kremlin-favored parties and candidates exactly matched their levels of support in opinion polls, it would have made no difference at all to the outcomes of presidential elections and very little to the distribution of power in the Duma.

Indeed, the forecasts have proven so accurate that some have wondered if this might indicate collusion between the pollsters and the Kremlin. While some organizations are closely connected to the presidential administration, the Levada Center is not, and, as noted in previous chapters, it was subject to a hostile takeover by the state apparently provoked by its determination to remain independent. There is no reason to doubt that polls it publishes faithfully represent the Center's best estimate of public opinion. To construct the forecasts, the pollsters used surveys for preceding weeks that asked people whether they planned to vote, and if so for whom. Voting preferences were then weighted by the respondent's likelihood of turning out.

Moreover, the election results fit well into the longer term patterns of support for different parties elicited in repeated polls over the years. Figure 8 shows the percentage of respondents who, in VCIOM/Levada Center polls, said that if a Duma election were held the next Sunday they would vote for, respectively, the Putin-connected Unity or United Russia parties; the Communist Party; and Russia's Choice or its successor, the Union of Right Forces. With the letters "CP," "U," "UR," and "RC," I indicate the share of the valid vote received by the Communist Party, Unity, United Russia, and Russia's Choice respectively in the actual elections held between 1993 and 2007. As can be seen, the electoral results for each party are close to the lines showing the trajectory of support reported by the survey respondents.[12]

---

[11] Where the pollsters do less well is in predicting turnout after 2000. They do a very good job in the 1990s, but the turnout predictions are quite a bit too high in 2003 and too low in 2007. However, it is not the case that when turnout was unexpectedly high, the Kremlin allies did better than expected.

The authorities' heavy-handed interventions hardly seem to have changed the results at all. According to Lev Gudkov, director of the Levada Center, turnout would have been more than 10 percentage points lower if the 2007 election had been run in a completely honest way. But the shares won by the various parties would have been about the same.[13] Another researcher tried to estimate how the results would change if one left out all regions with suspicious results. She concluded that United Russia would have received about 61 percent (instead of 64 percent) and the Communists would have got about one percentage point more.[14] Even the opposition seemed to accept that the widespread abuses had not made much of a difference. After the 2003 vote, Zyuganov accused the Kremlin of rigging the election. Nevertheless, his party's parallel tally of the votes had it winning just .03 percentage points more than it did in the official results.[15]

---

[12] I have interpolated values where data were missing. Where possible, I used fitted values from a regression of voting preferences for a given party on answers to another VCIOM/Levada question that asked with which party or group of parties respondents "sympathized." (For instance, I regressed the percent saying they would vote for the Communists on the percent saying they sympathized with the Communists and used fitted values to fill gaps in the data in the first series.) The series on sympathies of voters started later than the series on vote preferences but had fewer gaps. The correlations between sympathy for and vote preference for a given party were high. Where this was not possible, I simply interpolated linearly from the surrounding values in the series. Interpolations represent 41 percent of the data for the Communists and Russia's Choice and 23 percent for Unity/United Russia.

[13] "Itogi izbiratelnoy kampanii v Gosudarstvennuyu dumu pyatogo sozyva," Seminar at the Moscow Carnegie Center, 13 December 2007, http://monitoring.carnegie.ru/2007/12/analytics/seminar-2007-13-12.

[14] Ibid.

[15] "Communists say Duma vote was rigged," Gazeta.ru, 10 December, 2003.

# Figure 8.   Support for parties in opinion polls and official election results

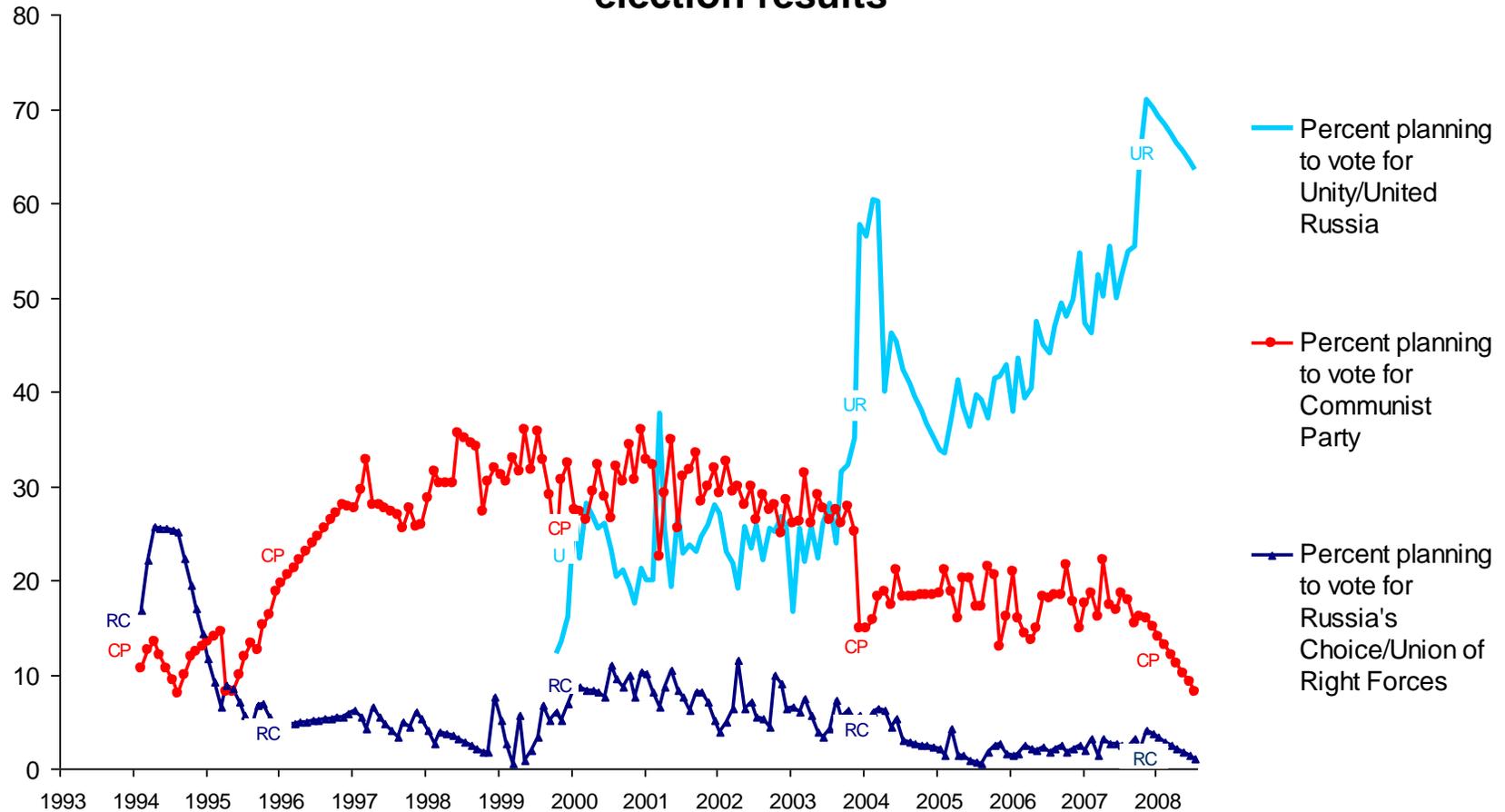

Source:  VCIOM/Levada Center polls, www.sofist.ru; vote plans measure percent indicating the given party when asked how they would vote if a Duma election were held the next Sunday, excluding those who would not vote or who were undecided; some values interpolated. U: percent for Unity; UR: percent for United Russia; CP: percent for Communist Party; RC: percent for Russia's Choice or Union of Right Forces--in Duma elections.

Manipulations or fraud are often said to have made a difference on three occasions. First, on the basis of certain unusual statistical features of the published results, a group of experts claimed that the 1993 referendum on the constitution had not received the required turnout of 50 percent of the electorate. The evidence they presented was far from unequivocal.[16] Second, although noone contested that Putin was the winner in 2000, some suggested that only irregularities allowed him to win in the first round, receiving more than 50 percent of the vote. Note, however, that the pollsters—FOM and VCIOM—had both predicted a vote share for Putin about as large as or slightly larger than the official result, in the range 53-55 percent. Finally, on the basis of their parallel vote count in 2003, the Communists claimed that the liberal parties Union of Right Forces and Yabloko both passed the five-percent threshold and should have been seated in the Duma. However, this claim was based on examining results in just 15 percent of the districts. Given that districts where the liberals were popular were likely to be ones where officials would be most open about publishing their protocols, it is very possible that the true results for the liberal parties were below those of the Communist count.

In sum, the evidence suggests that relatively minor electoral fraud occurred in the 1990s, and considerably more did after 2000. In the early 1990s, irregularities appear to have advantaged the Communists and, in 1999, an anti-Kremlin coalition of governors, rather than the Yeltsin administration and its allies. Since 2000, irregularities have clearly helped Putin and his associates. And yet, the manipulation and fraud do not seem to have changed the outcomes of elections much at all. Judging from what representative samples of Russians told pollsters in private interviews, the votes for the Communists and liberal reformers fell, and those for the parties connected to Putin rose over time, because most Russians came to genuinely support the pro-Kremlin parties and oppose the Communists and reformers.

Explaining nationwide trends

*Changing social composition of the electorate*

If electoral irregularities do not explain the voting trends, what does? Although voters switched back and forth quite fluidly early on, different parties did have distinct social bases of support. The Communist Party was particularly popular in rural areas, and among the old, less skilled workers, and those with lower incomes.[17] Zhirinovsky and the LDP appealed most to the young, and polled better among male than among female voters. The liberal reform parties showed strength in urban areas, and among the young and the highly educated.[18] Could it be that social and economic change had eroded some parties' bases, while enlarging those of others?

Focusing first on age, some wondered if the Communist vote was shrinking because, as one journalist put it, the party's elderly supporters were "steadily dying off."[19] Conversely, if the relative number of young adults was falling, that might explain the drop in support for the LDP and the liberal reformers. In fact, despite the decrease in Russia's total population, the ranks of both the elderly and young adults were swelling. The number of Russians aged over 60 rose from 22.5 million in 1989 to 24.5 million in 2008, while the number of Russians in their twenties grew from 22.3 million to 24.2 million.[20]

Perhaps what mattered was not age per se but having grown up under Stalin. It might be the passing of a generation of true-believers, socialized into Soviet beliefs during the country's totalitarian age, that was now shrinking Communist support. If this were the case, one would expect the oldest cohort to become less pro-Communist over time, as the true-believers died and were replaced by others who had come of age later. In fact, the opposite was the case. The proportion of those aged 60 and older who voted Communist was higher in 2007 (24 percent) than in 1993 (15 percent).[21] It was Communist voting in all the other age groups that fell drastically.[22]

Nor were there fewer Russians with higher education to vote for the reformers. On the contrary, the proportion of Russians aged over 14 with some higher education rose from 11 percent in 1989 to 16 percent in 2002.[23] The proportions living in urban and rural areas changed hardly at all. Nor was it the case that the old or young were dropping out, choosing not to vote in elections. In fact, the self-reported turnout rates for both the youngest and the oldest age groups increased by more than the average between 1993 and 2007.[24] It is true that turnout among the highly educated did not increase as much as for other groups—but this was because it started so high, just below 75 percent. In short, it does not appear that changes in the population's social characteristics can explain the trends in voting behavior.

*Economic conditions, presidential coattails, wars, campaigns*

What else might matter? In previous work, I found a close association between economic conditions, as perceived by the public, and the popularity of the incumbent president. We might expect a similar link between economic conditions and support for the main parties. Support for the party associated with the incumbent government should go up when the economy improves, while support for the opposition should increase when the economy deteriorates. Beyond such economic factors, one might expect voters' attitudes towards the incumbent president to influence their view of the party running on his coattails. High approval of Putin should translate into strong support for United Russia, while high approval of Yeltsin should boost the vote for Russia's Choice. In addition, voters might favor or oppose particular parties based on their positions on the two wars in

---

[20] Goskomstat RF, *Demografichesky yezhegodnik Rossii,* various years.

[21] These figures are the proportions of those that chose the Communists among those aged 60 and over who said they had voted, calculated from VCIOM post-election polls at http://wciom.ru/zh/print_q.php?s_id=481&q_id=36024&date=15.11.1993 and http://wciom.ru/zh/print_q.php?s_id=461&q_id=35547&date=09.12.2007.

[22] See also D. Roderick Kiewiet and Mikhail Myagkov, "Are the Communists dying out in Russia," Pasadena, CA: Caltech, 2002.

[23] Mezhgosudarstvenny statistichesky komitet Sodruzhestva Nezavisimykh gosudarstv, Statistika SNG. Statistichesky biulleten, Jun 15, 2005 .

[24] Calculated from VCIOM post-election polls.

Chechnya; in this case, as public opinion on the wars changed, so might the public's party preferences. Finally, one might expect the parties' performance during electoral campaigns to win or lose them voters.

To see which, if any, of these factors were important, I analyzed the responses to the VCIOM/Levada Center polls on voting preferences shown in Figure 8. Tests suggested that each of these series was fractionally integrated.[25] I therefore fractionally differenced the series.[26] I then ran error correction models, including variables to capture economic perceptions of the public (derived from VCIOM/Levada Center polls), levels of support for the incumbent president, whether or not a war in Chechnya was ongoing, and whether or not the month fell within a presidential campaign or its aftermath (I used a variable that took the value 1 for the six months preceding the presidential election and -1 for the six months following it). The economic perceptions and presidential support variables were themselves fractionally integrated in most cases, so I fractionally differenced these too, using the same technique as for the voting preference series. Where necessary to reduce autocorrelation to acceptable levels, I included one or two lags of the fractionally differenced dependent variable.

I tried including three economic perceptions variables. To measure retrospective evaluations of the national economy, I used the VCIOM/Levada Center survey question: "How would you assess Russia's present economic situation?" Respondents could answer "very good," "good," "in between," "bad," "very bad," or "don't know." I added the percentages saying "very good" or "good" and subtracted the shares saying "very bad" or "bad."[27] To measure retrospective evaluations of personal finances, I used the question: "How would you assess the current material situation of your family," which had the same choice of answers. Again, I subtracted the percentages saying "very bad" or "bad" from those saying "very good" or "good." For prospective evaluations of the national economy, I used the question: "What do you think awaits Russia in the economy in the coming several months?" Respondents chose between "a significant improvement of the situation," "some improvement of the situation," "some deterioration of the situation," "a significant deterioration of the situation," and "don't know." I subtracted the percentage anticipating deterioration from that anticipating improvement. No question was available for a comparable time period on prospective views of personal finances.

I also tried including dummy variables for particular events—most notably, for the month of December 2000, in which Putin restored the Soviet era music to the Russian national anthem, and December 2003, the last month of that year's Duma election campaign, when, as is clear from Figure 8, a major change occurred in support for both the Communists and United Russia. If this month was different in some important way, omitting the cause for this difference might bias the results for other variables.

---

[25] On fractional integration and appropriate methods to analyze fractionally integrated series, see, for instance, Janet M. Box-Steffensmeier and Andrew R. Tomlinson, "Fractional Integration Methods in Political Science," *Electoral Studies*, 19, 1, March 2000, pp.63-76.

[26] I used James Davidson's Time Series Modeling software to calculate *d* for each series, and averaged the results for bandwidths of 10, 20, and 30.

[27] This was generally available every second month, with some longer and some shorter gaps. To avoid irregular gaps in the data, I interpolated missing values linearly.

To review the results: first, I found no evidence that either of the Chechnya wars had a direct effect on Russians' party preferences. For the first Chechen war, I simply included a variable that took the value 1 in the first month of the first Chechen war, -1 in August, 1996, the month in which the Khasavyurt Accord was signed, and 0 in all other months. Dating the end of the second Chechen war was more complicated. I used a variable measuring the proportion of respondents in VCIOM/Levada Center polls that when asked what was currently occurring in Chechnya said that "war was continuing." This series appeared to be fractionally integrated, so I included both the lag of the level and the fractionally differenced value. As can be seen, these variables were not statistically significant for any of the three parties.

Second, there was evidence that Russians' perceptions of the state of the economy influenced party preferences. When Russians considered their personal finances to be in bad shape and when they thought national economic conditions were deteriorating, more

Table 3.  Explaining party preferences of voters, 1994-2008

| Dependent variable: | $F$ Communist Party | | $F$ Unity / United Russia | | $F$ Russia's Choice/ Union of Right Forces | |
|---|---|---|---|---|---|---|
| | (1) | (2) | (3) | (4) | (5) | (6) |
| Lag Communist Party | -.11*** (.04) | -.12*** (.04) | | | | |
| Lag Unity/United Russia | | | -.13 (.08) | -.26*** (.10) | | |
| Lag Russia's Choice / URF | | | | | .03 (.05) | -.30*** (.08) |
| Lag second Chechen war | -1.23 (1.25) | | -.58 (8.94) | | -1.78 (1.55) | |
| Lag current economy | -.02 (.03) | | .11 (.15) | .16* (.09) | -.03 (.04) | -.06*** (.02) |
| Lag current economy before May 1995 | | | | | | .25*** (.06) |
| Lag family finances | -.06 (.04) | -.08** (.03) | -.12 (.12) | | -.01 (.04) | |
| Lag Russia's ec. future | .05 (.03) | .03 (.02) | .06 (.06) | .08 (.08) | -.01 (.02) | .02 (.02) |
| Lag Yeltsin's approval rating | | | | | .57 (.39) | .69** (.30) |
| Lag Yeltsin's approval rating before May 1995 | | | | | | 6.92*** (1.50) |
| Lag Putin's approval rating | | | 1.51 (2.04) | 1.61 (2.08) | .61* (.32) | .54** (.21) |
| Lag Presidential campaign | | | | -7.58 (9.94) | | |
| Lag presidential campaign * Lag Putin's approval rating | | | | .96 (1.48) | | |
| $F$ second Chechen war | -1.33 | | 5.20 | | 4.99 | |

| | (1) | (2) | (3) | (4) | (5) | (6) |
|---|---|---|---|---|---|---|
| | (9.87) | | (16.41) | | (4.37) | |
| F current economy | -.14** | -.15** | .26 | .21 | .02 | |
| | (.07) | (.07) | (.18) | (.15) | (.05) | |
| F family finances | -.04 | | -.09 | | -.04 | |
| | (.05) | | (.21) | | (.08) | |
| F Russia's ec. future | .12** | .10* | -.06 | | -.07 | -.05 |
| | (.05) | (.05) | (.12) | | (.05) | (.03) |
| F Yeltsin's approval rating | | | | | 5.13*** | 2.05* |
| | | | | | (1.89) | (1.07) |
| F Yeltsin's approval rating before May 1995 | | | | | | 13.73*** |
| | | | | | | (3.85) |
| F Putin's approval rating | | | 1.42 | -1.13 | .61 | .42 |
| | | | (1.20) | (2.20) | (.51) | (.34) |
| Lag presidential campaign * F Putin's approval rating | | | | 5.31** | | |
| | | | | (2.50) | | |
| First Chechen war | .41 | | | | -1.17 | |
| | (.48) | | | | (.97) | |
| Soviet anthem | 4.90*** | 4.95*** | 1.71 | 3.32*** | 2.70*** | 2.58*** |
| | (.64) | (.57) | (1.36) | (1.09) | (.35) | (.26) |

Table 3 continued

| | $F$ Communist Party | | $F$ Unity / United Russia | | $F$ Russia's Choice/ Union of Right Forces | |
|---|---|---|---|---|---|---|
| December 2003 | -9.89*** (.60) | -9.69*** (.61) | 19.58*** (1.40) | 21.39*** (1.27) | .68* (.40) | .49** (.24) |
| Presidential campaign | -.83** (.36) | -.81** (.34) | 3.46*** (1.09) | | -.04 (.21) | |
| Lag dependent variable | -.37*** (.09) | -.35*** (.09) | -.25** (.10) | | | |
| Second lag dependent variable | -.26*** (.09) | -.24*** (.09) | | | | |
| Constant | .37 (.88) | .15 (.76) | -2.77 (16.10) | 6.87 (15.80) | -4.26** (2.11) | -4.62*** (1.51) |
| $R^2$ | .3219 | .3124 | .4213 | .3493 | .1143 | .3315 |
| Ljung-Box (residuals), Q(12) | 10.44 (p = .58) | 9.82 (p = .63) | 11.75 (p = .47) | 5.12 (p = .95) | 15.16 (p = .23) | 11.33 (p = .50) |
| N | 168 | 168 | 101 | 102 | 170 | 170 |

$F$  fractionally differenced series. Second models for each dependent variable include interaction terms to distinguish particular time periods, and exclude some variables that have very low statistical significance.

people said they would vote for the Communists (model 2). (A little surprisingly, when expectations for the national economy deteriorated, fewer planned to vote for the Communists.) Economic conditions had different implications for Russia's Choice depending on whether or not it was the main party of government. While it was still the party of power, the popularity of Russia's Choice was higher when the economy was doing better. But after Russia's Choice surrendered this role to OHIR in May 1995, becoming if not an opposition party at least the advocate of an alternative approach to reform, its appeal was greatest when economic conditions were bad.[28] Later on, United Russia also did better when the economy was perceived to be strong, although the evidence on this is less clear (the coefficient on the lag of views of the current economy was positive and marginally significant in model 4).

The fortunes of both United Russia and Russia's Choice were clearly tied to the popularity of the presidents with which they were associated. As one might expect, a rise in Putin's rating was associated with an increase in the proportion of Russians that planned to vote for Unity/United Russia. What is interesting, however, is that such convergence *only occurred during the presidential*

---

[28] I include variables for the lag of evaluations of the current economy and for the same thing before May 1995, when OHIR was founded. Thus, the estimated effect of this before May 1995 is .25 - .06 = .19, while the estimated effect after May 1995 is -.06 (see model 6). Along with the negative coefficient on the lagged percent planning to vote for RC/URF, these imply a positive long-run relationship between evaluations of the current economy and plans to vote for RC/URF before May 1995 and a negative long-run relationship after May 1995.

*election campaigns*. In fact, a rise in Putin's rating at other times was associated with, if anything, a fall in United Russia's support (though this is not statistically significant). But in the heat of the electoral season, the party's ratings would leap upward closer to Putin's; then after the election was over, some of the party's support would quickly dissipate.[29]

To anyone who observed these campaigns, this makes a lot of sense. Unity's and United Russia's advertising sought to get across just one, quintessential point about the party—that it was the party of Putin. Other than that, it had almost no identity. In 2007, the year in which Putin agreed to head United Russia's party list, some of its electoral posters did not even mention the party's name. One, for instance, just showed a photograph of Putin, alongside the caption: "2 December—Elections—Of Putin."[30]

Throughout Yeltsin's two terms, the popularity of Russia's Choice was linked to that of the president. But the link was much stronger before OHIR took over as the main pro-government party.[31] During Putin's terms in office, Russia's Choice's successor, the Union of Right Forces, did better when Putin's approval was high.

Besides economic perceptions and the president's popularity, two other phenomena shaped the pattern of party support. First, in December 2000, Putin restored the Soviet-era music to the Russian national anthem. This act appears to have polarized the political community. The ratings of Unity, the Communists, and Russia's Choice *all* jumped by several percentage points. The debate over this issue had been divisive, with both the Communists and President Putin strongly supporting restoring the traditional music, and former President Yeltsin breaking his post-retirement silence to criticize the decision. Liberals opposed it, with the Yabloko faction complaining that the restoration "deepened the schism in society."[32] It appears that the debate enabled all the factions to rally previously netural citizens behind them.

Second, , in the final month of the 2003 Duma campaign, a major reshuffling of support occurred. The proportion saying they would vote for United Russia rose by around 20 percentage points, while the proportion endorsing the Communists fell by around 10 points.[33] I cannot say for sure what caused this. There are several possibilities. The Kremlin's political operatives had unleashed a ferocious onslaught against the party, accusing Zyuganov and his allies of selling out to the oligarchs, one of whom Putin had just jailed. It turned out that the Communists had at least four dollar millionaires on their party list, two of them former executives from Khodorkovsky's company.[34] Then, in an embarrassing leak, a Communist insider claimed that the party had been in contact with the even more scandalous tycoon, Boris Berezovsky. The Communist nominee for governor of Omsk said that he and the party's ideology secretary had met with Berezovsky in London in 2002 to importune him for money on the party's behalf.[35]

---

[29] See model 4, coefficient on "lag presidential campaign * FΔPutin's approval rating"; recall that the presidential campaign variable took the value 1 in the six months before a presidential election and -1 in the six months after the election.

[30] I am grateful to Konstantin Sonin for the photograh of this poster.

[31] There is evidence of both long-run equilibrium relationships and short-run adjustment effects; see the coefficients on the lag of Yeltsin approval and of this for just the period before May 1995, and the coefficients for the fractional difference of Yeltsin approval and for this before May 1995, in model 6.

[32] BBC, "Duma approves Soviet anthem," 8 December 2000, http://news.bbc.co.uk/2/hi/europe/1060975.stm.

[33] These are estimates from regressions controlling for economic performance and/or presidential approval.

At the same time, in order to steal votes from the Communists, the Kremlin had sponsored the creation of a new party called Fatherland, which combined nationalist and social democratic appeals and was led by a left-wing economist called Sergey Glazyev along with an outspoken nationalist, Dmitry Rogozin.[36] The last straw for Zyuganov's supporters may have been his decision to drop out from the presidential election, apparently under pressure from the administration, and to send a lacklustre associate in his place.[37] Bewildered and disappointed by this, the party faithful may well have concluded that he had been coopted by the regime. Still, December 2003 only accelerated the downward trend that had begun around early 2000, and which can be well-explained by just economic perceptions.

Figure 9A shows the actual share of voters that said they would vote for the Communist Party if Duma elections were held the next Sunday, along with the share that could be predicted using just economic perceptions variables (that is using the estimates

---

[34] Vitaly Ivanov, Anfisa Voronina, and Anna Nikolayeva, "Krasny millionery," *Vedomosti*, October 15, 2003. Among the Communist millionaires was Gennady Semigin, who for a while served as vice-speaker of the Duma and had assets worth $16 million in 2003. The two former Yukos executives were Aleksey Kondaurov and Sergey Muravlyenko (with a fortune of around $47 million).

[35] Leonid Mayevsky interviewed by Vitaly Trubetskoy, "Berezovsky zavyot na barrikady," *Vesti nedeli*, Rossiya TV channel, November 2, 2003, http://www.vesti7.ru/archive/news?id=3271.

[36] On Kremlin involvement in the party's creation, see Baker and Glasser, *Kremlin Rising,* pp.297-304.

[37] For the Kremlin pressures, see Baker and Glasser, *Kremlin Rising*, p.318.

**Figure 9A. Share of respondents who would vote for the Communist Party, and share predicted using just economic perceptions**

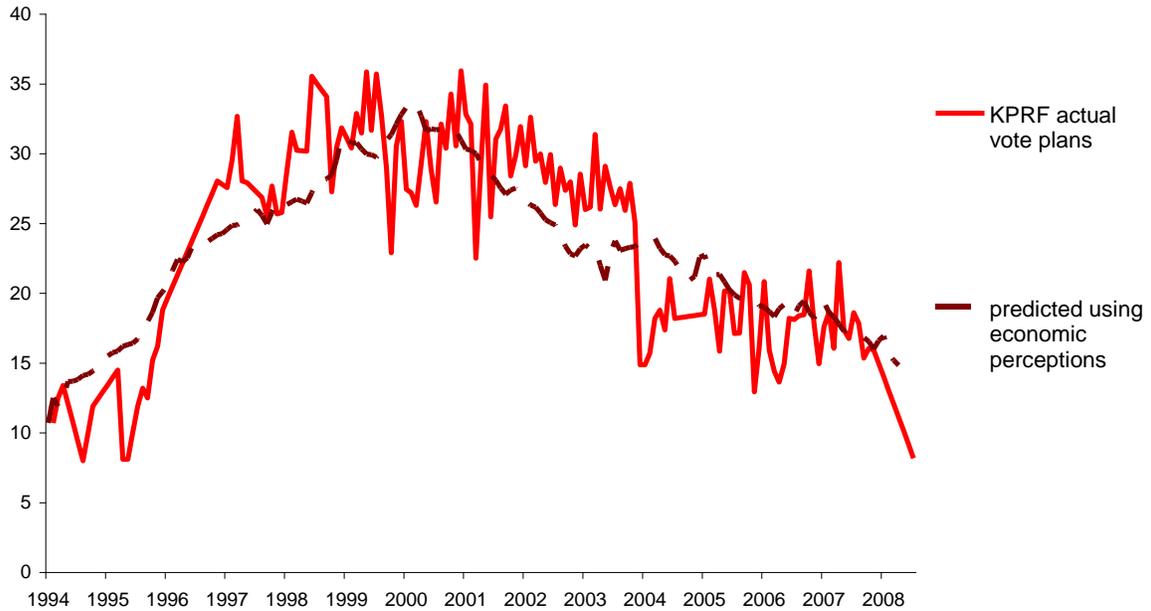

Source: VCIOM/Levada Center surveys and author's calculations.

**Figure 9B. Share of respondents who would vote for Unity/UR, and share predicted using just economic perceptions, Putin's rating, and campaign period**

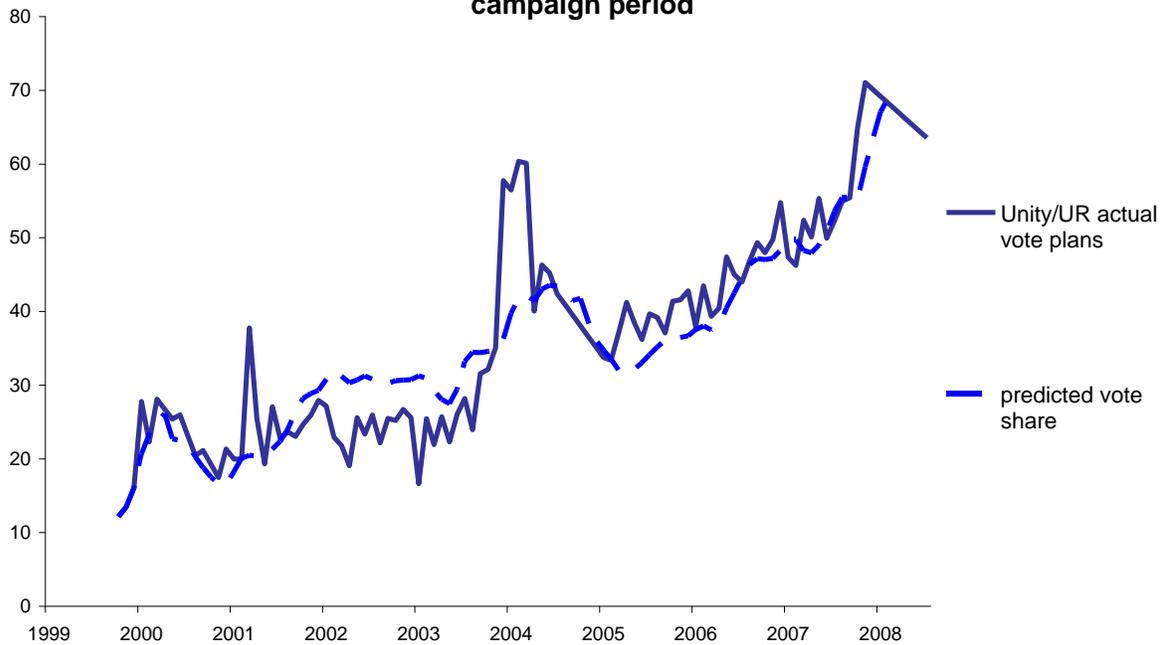

Source: VCIOM/Levada Center surveys and author's calculations.

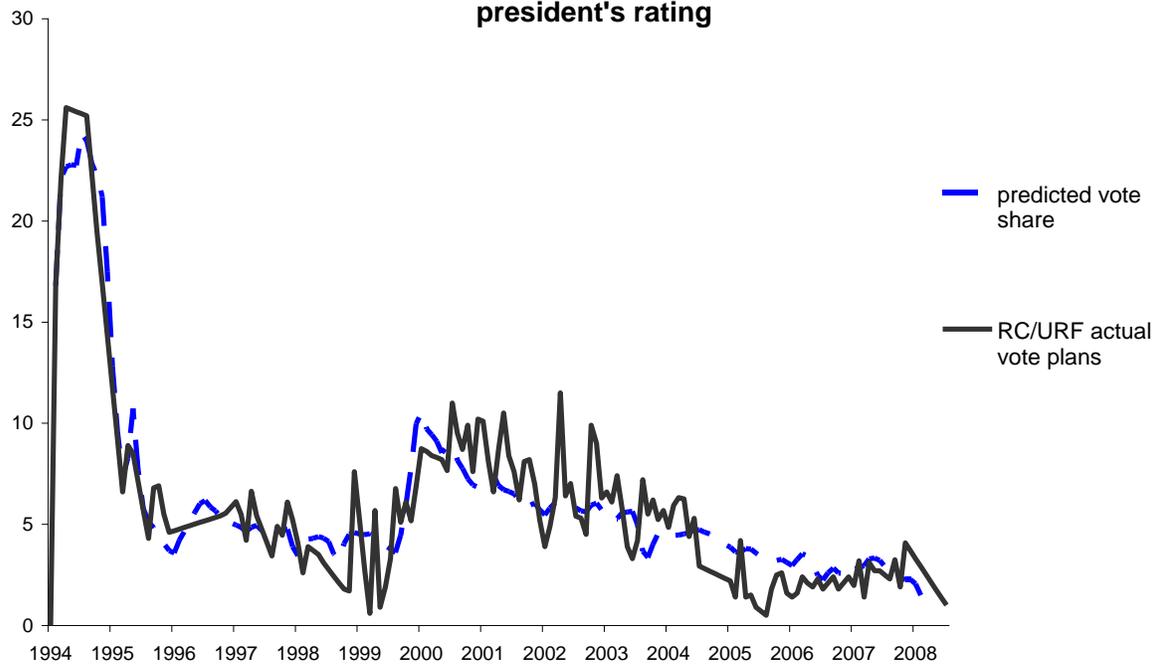

**Figure 9C. Share of Respondents who would vote for Russia's Choice/URF, and share predicted using economic perceptions and president's rating**

- - - predicted vote share

—— RC/URF actual vote plans

Source: VCIOM/Levada Center surveys and author's calculations.

from model 2 in Table 3, but leaving out the Soviet anthem, December 2003, and presidential campaign variables). Since working backwards from a fractionally differenced series to produce predicted values is a laborious process, I have instead used as an approximation predictions based on an otherwise identical regression in which the dependent variables are in first differences. As can be seen, the predictions based on economic variables fit the trend in pro-Communist opinion very well. Figure 9B shows the share planning to vote for Unity/United Russia and the share that would be predicted from Model 4, in first differenced form, leaving out the effects of the Soviet anthem and December 2003. Thus, it shows how well just economic perceptions, Putin's popularity, and the electoral campaign season can predict voting plans. The fit is good. Finally, Figure 9C shows the actual share of respondents who said they would vote for Russia's Choice/Union of Right Forces and the share predicted using economic perceptions and presidential popularity, and distinguishing between the pre- and post-May 1995 periods. All in all, while there is some variation around the trends, economic perceptions and presidential popularity—sometimes focusing on just the campaign season—do a very good job of predicting the trends in party preferences for these three parties.

The geography of voting

The power of economic variables to account for trends in voting behavior was also evident when I ran panel regressions to explain the regional vote levels for the Kremlin-associated candidates in the four presidential elections between 1996 and 2008 (that is, Yeltsin (1996), Putin (2000, 2004), and Medvedev (2008)) (see Table 5). To measure economic effects, I included variables for the level of real income per capita in the region two years before the election and the percentage change in real income per capita in the subsequent year (for example, in the case of the 1996 election, using the 1994 level and the 1994-5 change).[38] I also included a variable for the natural log of the regional wage arrears per worker as a percentage of the average wage. For 1996, wage arrears were measured as of January 1996, the first month for which data were available. For subsequent elections, I used the average wage arrears per worker as a percentage of the average wage for the preceding year (1999, 2003, 2007).

---

[38] I deflated using the cost of Goskomstat's "fixed basket of consumer goods and services for making interregional comparisons of purchasing power," as of December of each year. This was available only after 2002, so I extended this backwards using the December-to-December change in the consumer price index.

Table 3.  Regional Vote for Incumbents, presidential elections, panel regressions

| | OLS, PCSE | OLS, PCSE, period dummies | OLS, PCSE, period dummies | Regional fixed effects, clustered SEs | Regional fixed effects, clustered SEs, period dummies |
|---|---|---|---|---|---|
| | (1) | (2) | (3) | (4) | (5) |
| *Controls* | | | | | |
| Vote for incumbent previous election | -.12 (.19) | .32* (.17) | .32* (.17) | | |
| Republic | 6.49*** (1.43) | 5.25*** (1.29) | 5.33*** (1.33) | | |
| AO | 5.51* (2.90) | .79 (2.73) | .96 (2.96) | | |
| *Economic and fiscal factors* | | | | | |
| Real income per capita, 2 year lag | .27 (2.03) | .70 (1.58) | .62 (1.61) | -7.98** (3.51) | -5.52* (2.99) |
| Change in real income per capita, previous 2 years (%) | .69*** (.19) | .06 (.04) | .06 (.04) | .79*** (.07) | .07 (.08) |
| Ln real wage arrears as % of monthly wages, previous year | -.90 (1.04) | -1.32*** (.41) | -1.33*** (.42) | -.81 (.81) | -1.12** (.55) |
| Real regional budget expenditures, 2 year lag[a] | 4.60 (3.42) | 3.42* (1.86) | 3.15* (1.85) | 6.25 (4.65) | 2.86 (2.41) |
| Change in real regional expenditures, previous 2 years (%) | .16** (.07) | .07*** (.03) | .07*** (.03) | .14** (.06) | .08* (.04) |
| Share of central transfers in regional spending (%) | .12** (.06) | .09*** (.03) | .09*** (.03) | .21** (.10) | .02 (.07) |
| *Motives of regional leaders* | | | | | |
| Governor associated with Kremlin party | 5.54** (2.19) | .70 (1.14) | .71 (1.16) | 6.24*** (1.10) | .78 (.82) |
| *Capacity to manipulate electorate* | | | | | |
| Percent of local electoral commission members state appointees, late 1990s; all elections | .03 (.05) | .06* (.04) | | | |
| State appointees, 2000 election | | | .07** (.03) | | |
| State appointees, 2004 election | | | .10*** (.03) | | |

| | (1) | (2) | (3) | (4) | (5) |
|---|---|---|---|---|---|
| State appointees, 2008 election | | | .13*** | | |
| | | | (.04) | | |
| *Period dummies* | | | | | |
| 2000 | | 27.16*** | 26.35*** | | 19.62*** |
| | | (3.39) | (3.56) | | (1.14) |
| 2004 | | 32.91*** | 31.97*** | | 35.40*** |
| | | (1.73) | (1.69) | | (2.55) |
| 2008 | | 20.53*** | 19.38*** | | 32.81*** |
| | | (3.56) | (3.34) | | (4.10) |
| Constant | 56.14*** | 16.30* | 17.43* | 57.20*** | 41.26*** |
| | (9.84) | (9.69) | (10.11) | (6.29) | (4.39) |
| $R^2$ | .6285 | .8505 | .8517 | .5807 | .7721 |
| N | 318 | 318 | 318 | 319 | 319 |

* $p < .10$, ** $p < .05$, *** $p < .01$.

A related set of variables measure regional budget spending.[39] I included measures of real regional budget expenditures for two years before the election (i.e. 1994, 1998, 2002, and 2006) and the percentage change in expenditures between this and the subsequent year. I deflated using the deflator based on Goskomstat's basket of consumer goods and services discussed in the previous paragraph. It might be that more fiscally dependent regions would vote for the incumbents out of fear that "incorrect" voting might lead to a cutoff of subsidies. I therefore controlled for the share of the region's budget expenditures, as of the previous year, that was paid for by transfers.

I include two variables designed to capture effects of electoral manipulation or even fraud. First, I include a dummy for whether the region's governor was associated with the leading pro-Kremlin party. In 1995, it was a dummy for whether the governor ran on the OHIR list in the parliamentary election, and in 2003 and 2007 it was a dummy for running on the United Russia list. In 1999, the variable was a dummy for whether the governor was among those that signed a declaration in support of the Unity Bloc on September 29, 1999. A second variable measures the proportion of the members of local electoral commissions in the region (as of 1997-8) that were state employees. This is a rough-and-ready indicator for the extent to which the administration of elections in the region was controlled by the executive branch. As the general trend towards greater manipulation intensified, one might expect the effects to be most noticeable in such regions. Finally, I also control for republic and autonomous okrug or oblast status, since journalistic reports of electoral irregularities were particularly common in the non-Russian ethnic regions and these also might have different trends in voting preferences associated with their ethnic minority populations.

I estimate these regressions using both OLS with panel-corrected standard errors and fixed effects models. In the OLS regressions, I include the lagged value of the dependent variable—the vote share for the incumbent in the previous presidential election (I treat Yeltsin as the incumbent in 1991)—since recent changes in the economic and fiscal variables are likely to affect change in parties' vote shares more than their absolute levels. Including the lagged dependent variable is important here also because of likely autocorrelation. However, I do not include lags of the dependent variable in the fixed effects models, as this can introduce endogeneity. I try both with and without dummies for the period—i.e. the election year—with the excluded category being 1996. The advantage of including period dummies is that it makes the regressions a tougher test of the significance of other factors and reduces the chances of picking up a spurious common time trend between explanatory variables and election results. The disadvantage is that the period dummies absorb a large part of the interesting variation; the regional deviations from the average effect may be small and the regressions therefore unable to pick up important causal processes that have a strong common element that is shared across regions. As we will see, this makes a difference for some of the results. In the fixed effects regressions, I cluster the standard errors by region.

First, note that there is consistent evidence of the importance of regional budget expenditure. In regions where the government increased its spending more (or decreased it less) in the years leading up to a presidential election, the vote for the incumbent tended to increase relatively more compared to the previous vote. In most models, it also appeared that regions that were more fiscally dependent tended to have larger increases in the vote share of the incumbent. There was some evidence of economic influences. Where wage arrears were larger, the vote for the central incumbent tended to be lower, other things equal. Judging by the regressions without period dummies, faster growth in real incomes also correlated with relatively larger increases in votes for the incumbent. These effects were not statistically significant in the regressions with period dummies. In fact, this is exactly what one should expect. If rapid income growth in a given region matches a nationwide trend, then it is reasonable for the voters to credit this to policies of the central government and vote accordingly. However, if incomes are growing fast within a given region, but not elsewhere, then voters are more likely to attribute this to local circumstances or regional policy and not to central policies. It would be odd, in this case, for local voters to reward the central incumbents. Yet, in regressions that include period dummies, the effect of any nationwide trend in incomes would be picked up by the dummies, and only the impact of regional divergences would be captured by the income variable. (Regressions without the period dummies might be considered problematic because the rates of income growth in different regions are not independent. However, the panel corrected standard errors in column 1 adjust for contemporaneous correlation.) In short, the data are very consistent with the idea that rising incomes led to greater support for the Kremlin's presidential candidate, although one cannot be completely sure that some other nationwide trend correlated with the change in real incomes is not really responsible.

The representation of state employees on local electoral commissions is just marginally significant in the basic regressions. But this turns out to be because the effect varied greatly over time. In 1996, more state employees on the electoral commissions were not associated with any advantage for the incumbent. However, as column 3 shows, if the effect is broken down by election year, we see strong and increasing effects in 2000-08 of this indicator of executive branch dominance of electoral administraiton. It also appears that when a region's governor was associated with the pro-Kremlin party, the vote for the central incumbent rose faster. The number of governors associated with the Kremlin party increased over time, along with support for the central incumbents, and there is not enough variation in the variable to deduce much from the deviations from the trend. Thus, we cannot be quite as sure of this result. Causation might also go the opposite way: that is, governors in regions where the central incumbent is popular might choose to side with the pro-Kremlin party, rather than pro-Kremlin governors manipulating the vote. Finally, note the strong effect of the ethnic republic dummy, even in regressions that control for the previous election result. In the republics, other things equal, the vote for the incumbent was rising faster than in the non-ethnic regions.

*The ethnic regions*

The ethnic regions' affinity for Putin is one of the strangest phenomena in recent Russian politics. On the face of it, this seems bizarre. After all, Putin came to the Kremlin with the explicit aim of clawing back the privileges and prerogatives the regional leaders had negotiated out of Yeltsin in the 1990s. He abrogated the power-sharing agreements his predecessor had signed with the regional capitals. He forced richer regions to pay more into the central budget. Between 1999 and 2003, Tatarstan's remittances rose from 13 to 49 percent of locally collected taxes. The oil-rich Khanti-Mansiiskiy autonomous district's remittances went from 40 to 72 percent.[40] He sent troops back into Chechnya to crush its separatist warlords, and began simply writing some of the smaller autonomous okrugs off the map. The Komi-Permyaksky and Ust-Orda districts were merged into the surrounding predominantly Russian regions.[41]

And yet, these same ethnic regions turned in ever higher votes for the centralizer in the Kremlin. Did this reflect something about the specific attitudes of members of non-Russian ethnicities? For the most part, the evidence suggests it did not. Using the Russian Longitudinal Monitoring Survey (RLMS), I compared the proportions approving of Putin among respondents of self-reported Russian and non-Russian nationalities. I found that what mattered was mostly not the nationality of the individual but where he or she lived. Among those living in areas where non-Russians predominated, approval of Putin's performance was much higher than elsewhere. And this was true among both ethnic Russians and non-Russians in those places. In locations where ethnic Russians predominated, approval of Putin's performance was close to nationwide trends—among both ethnic Russians and ethnic non-Russians (see Table 4). Extremely high support for Putin appears to be an attribute of place rather than individual identity.

What was it about the ethnic regions that produced such high levels of support for Putin? Conceivable explanations range from the innocent to the cynical. Some of the ethnic regions are among the most undeveloped in Russia and dependent on the central authorities for fiscal assistance. This might cause local citizens to vote for central incumbents out of fear these subsidies might be withdrawn. As one Dagestani journalist put it, the republic's reliance on federal transfers for about 90 percent of its budget meant that "Dagestanis are obliged to bet only on the favorite in Russia's presidential elections."[42]

Table 4: Approval of President Putin, by type of site, nationality and religion, late 2003

| | Predominantly Russian sites | | | | |
| --- | --- | --- | --- | --- | --- |
| | All respondents | Russian | Non-Russian | Muslim | Orthodox |
| Completely approve or approve | 57.1 | 55.9 | 61.1 | 64.4 | 57.7 |
| Approve of some things, not others | 35.9 | 36.7 | 33.3 | 30.1 | 35.9 |
| Disapprove or completely disapprove | 7.1 | 7.5 | 5.7 | 5.6 | 6.4 |

[40] On the other hand, a few poorer republics such as Altai Republic got to keep a larger share of local revenues. These figures are based on author's calculations, from Goskomstat Rossii statistics and Ministry of Finance reports on regional budget execution.

[41] Danielle Lussier, "Putin Continues Extending Vertical of Power," *Russian Regional Report*, 8, 2, 3 Feb. 2003

[42] Abdullaev, Nabi. 2000. "Dagestan's Leadership is Happy with Putin," *Prism*, Jamestown Foundation, 6, 4, April

| | All respondents | Russian | Non-Russian | Muslim | Orthodox |
|---|---|---|---|---|---|
| N | 6,983 | 5,867 | 706 | 216 | 5,720 |
| Mean (completely approve = 1, completely disapprove = 5) | 2.4 | 2.4 | 2.3 | 2.3 | 2.4 |
| *Predominantly non-Russian sites* | | | | | |
| | *All respondents* | *Russian* | *Non-Russian* | *Muslim* | *Orthodox* |
| Completely approve or approve | 74.3 | 72.7 | 74.7 | 75.3 | 72.9 |
| Approve of some things, not others | 19.4 | 27.3 | 18.8 | 16.3 | 26.0 |
| Disapprove or completely disapprove | 6.4 | 0 | 6.5 | 8.4 | 1.0 |
| N | 439 | 22 | 384 | 320 | 96 |
| **Mean (completely approve = 1, completely disapprove = 5)** | 1.9 | 2.0 | 1.9 | 1.8 | 2.1 |

Source: Author's calculations from RLMS Round 12.

Less innocently, observers have suggested reasons why ethnic regions might prove more electorally "manageable". The ethnic units tend to be more rural than the non-ethnic units: as of 2003, the average rural population share in the republics and autonomous districts was 43 percent, compared to 27 percent in the other regions. Rural districts are thought to be easier for the local authorities to control. This might be because populations are more traditional and less well-educated or because individuals are more dependent on the collective farm infrastructure, and so more subject to manipulation by the farm director. Farm directors, in turn, are often dependent on the regional governor for maintaining local infrastructure or helping to obtain federal subsidies. And it is easier to detect how the residents of an isolated farm community voted than, say, the workers in an urban factory, who may vote where they live in different corners of the city.[43]

Another possibility is that the solidarity and social networks within a minority ethnic community make organizing political machines easier.[44] In the ethnic regions, mutual advancement networks had often emerged earlier to take advantage of the Soviet affirmative action for members of the titular nationalities. A regional leader's ability to control local political affairs might be expected to increase with his time in office, and the autonomy given ethnic republics might enable their leaders to entrench themselves in office. At the same time, the autonomy given regional leaders in the republics might make it easier for them to control local officials—and local electoral commissions—enabling them to manage the vote better.

---

[43] Henry Hale, "Explaining Machine Politics in Russia's Regions: Economy, Ethnicity, and Legacy," *Post-Soviet Affairs*, 2003, 19, 3, pp. 228–263.

[44] Hale, "Explaining Machine Politics."

To try to better understand the ethnic effects, I examine the 2004 presidential election results a little more closely in Table 5. The goal is to see what variables, if added to the regression, reduce the size of the ethnic republic and autonomous okrug effects. Notice that these regressions aim to explain level effects rather than changes. I do not include a lagged dependent variable, so those variables—such as change in real incomes—that helped explain change in the vote share for the incumbents will be less powerful.

The ethnic effect was not caused by the more rural character of the average ethnic region. Controlling for the degree of urbanization or the share of agricultural workers in the workforce made very little difference to the coefficients on ethnic status. Among non-republics, the more agricultural regions had lower votes for Putin, but among the republics, agriculture had no effect.[45]

There was no evidence that regions where the governor had been in office for longer had higher pro-Putin votes. In any case, the average time in office, as of 2003, of the incumbent governor was almost exactly the same in the ethnic as in the non-ethnic regions—six years. There was indirect evidence that when the regional government had greater control over local administration—and in particular the administration of elections—this led to a higher pro-Putin vote. The variable measuring the percentage of members of the local electoral commissions that were state or local government employees was positively associated with the Putin vote. This may account for a small

---

[45] The effect of agriculture in the republics is equal to the coefficient on agricultural employment plus the coefficient on agricultural employment in the republics;  in model 5 this is -.47 + .48 ≈ 0.

Table 5.   Regional Vote for Putin, 2004

| | (1) | (2) | (3) | (4) | (5) |
|---|---|---|---|---|---|
| Republic | 12.81*** | 9.00*** | 8.94*** | 2.48 | -3.67 |
| | (2.51) | (2.31) | (2.26) | (2.16) | (3.81) |
| AO | 11.54*** | 8.14** | 7.49** | 6.95** | 6.56** |
| | (1.83) | (3.89) | (3.27) | (2.99) | (3.11) |
| *Economic and fiscal controls* | | | | | |
| Real income per capita 2002 | | 1.61 | 1.12 | -.10 | .36 |
| | | (2.47) | (2.60) | (2.42) | (2.37) |
| Change in real income per capita, 2002-03 (%) | | .06 | .06 | -.01 | -.00 |
| | | (.10) | (.09) | (.07) | (.07) |
| Ln of real wage arrears as a percent of monthly wages of workers, 1999 | | -.91 | -.49 | -.17 | .08 |
| | | (.73) | (.76) | (.53) | (.47) |
| Real regional budget expenditures, 2002[a] | | 2.24 | 1.94 | 2.01 | 1.82 |
| | | (1.54) | (1.22) | (1.26) | (1.28) |
| Change in real regional budget expenditures, 2002-03 (%) | | .11** | .12*** | .04 | .04 |
| | | (.04) | (.04) | (.05) | (.04) |
| Share of central transfers in regional spending (%) | | .15** | .17** | .13** | .11** |
| | | (.07) | (.06) | (.05) | (.05) |
| *Motives of regional leaders* | | | | | |
| Governor on United Russia list, 2003 | | 4.37** | 3.71** | 3.65** | 3.12** |
| | | (1.94) | (1.74) | (1.50) | (1.47) |
| *Capacity to manipulate electorate* | | | | | |
| Percent of local electoral commission members state appointees, late 1990s | | .11* | .11* | .11** | .09* |
| | | (.07) | (.06) | (.05) | (.05) |
| Employment in agriculture, % | | | -.30** | -.37*** | -.47*** |
| | | | (.13) | (.12) | (.12) |
| Employment in agriculture, republics, % | | | | | .48** |
| | | | | | (.19) |
| Ethnic region chief executive of titular nationality | | | | 11.89*** | 10.68*** |
| | | | | (2.71) | (2.31) |
| Year chief executive took office | | | | | -.05 |
| | | | | | (.16) |
| *Other controls* | | | | | |
| Crime rate, 2003 | | | -.004** | -.003** | -.003** |
| | | | (.001) | (.001) | (.001) |
| Constant | 67.86*** | 59.68*** | 69.70*** | 71.75*** | 181.79 |
| | (.65) | (4.66) | (4.52) | (4.23) | (318.39) |
| $R^2$ | .4305 | .5283 | .5816 | .6820 | .7054 |
| N | 89 | 83 | 83 | 83 | 83 |

Heteroskedasticity corrected standard errors in brackets. * p < .10, ** p < .05, *** p < .01.

part of the ethnic region difference.[46] Governors who had run on the United Russia list in 2003 also tended to have votes for Putin in their region a few points higher than elsewhere, but this cannot explain the ethnic region effect—controlling for this increases the coefficients on republic and autonomous okrug status.

Two variables did help to explain the ethnic region effect. First, it appeared to be crucial that the region's leader was himself of the titular nationality that gave the region its name. In late 2003, this was the case in 14 of the 31 ethnic regions In ethnic regions where the governor was Russian or Ukrainian, the vote for Putin averaged 75 percent. But in those with governors of the locally dominant nationality, his average vote was 87 percent. Based on the model 4 regression, republics with a non-titular leader had on average only 2.5 percentage points higher votes for Putin than non-ethnic regions, other things equal. Second, ethnic regions had higher votes for Putin in part because they tended to be more fiscally dependent on the central government, and apparently believed that higher pro-Putin votes were necessary to obtain continued subsidies. Other things equal, a republic like Ingushetia, which in 2003 relied on federal aid to finance 82 percent of its budget, had a vote for Putin about eight percentage points higher than a region like the City of Moscow, where federal aid paid for only about seven percent of spending.

Why having a leader of the titular nationality should make such a difference for the public's voting is not clear. Of course, it could be that some other factor determines both which regions have a titular head and which have a high vote for Putin. But the result held controlling for a variety of other factors. For instance, predominantly Muslim ethnic regions and those in the South were more likely to have a leader of the titular nationality. But neither of these facts could explain the result: having a leader of the titular nationality was much more significantly associated with high Putin voting among the ethnic republics than Muslim population or southern location. The most plausible explanation is that the ethnic solidarity and perhaps the cultural networks uniting a titular nationality leader with the local administrative elite made it easier to mobilize the population to vote in a certain way or to falsify the election without fear of exposure.[47]

## Conclusion

Analyzed systematically, the results of elections and opinion polls on party preferences in Russia in the first 18 years of its postcommunist politics show some quite intelligible patterns. Two phenomena have been particularly important. The dominant influence on voting preferences over time has been change in the economy. Changes in perceived economic performance—both directly and indirectly, through their effects on the popularity of incumbent presidents—have apparently driven change in the levels of nationwide support for the Communists and for pro-regime parties such as Russia's Choice or United Russia. Economic and fiscal factors—especially the level of wage arrears, the depressed state of agriculture, and changing levels of regional government spending—help to explain the changing geographical patterns of support for Kremlin-backed candidates in presidential elections.

[46] Controlling for the share of state employees on the local electoral commissions reduces the coefficient on republic status in model 1 from 12.81 to 10.63; the coefficient on autonomous okrug falls to 10.73. However, the average share of state employees on the local electoral commissions was similar in the ethnic and non-ethnic regions.

[47] On ethnic machine politics in Russia, see Hale, "Explaining Machine Politics."

A second factor—the increasing manipulation or even outright falsification of election results—has been important for shaping the geographical pattern of electoral results, but has apparently not been nearly as important as many people think in determining the nationwide results for particular candidates and parties. At least that is the conclusion one reaches when one compares election results to the results of professionally conducted opinion polls. Manipulations during elections have not changed the results much from what recent polls predicted. However, some kind of manipulation is the likeliest explanation for the distinctive pattern of voting within the ethnic regions, which have turned into the Kremlin's most enthusiastic cheer-leaders. And the apparent importance of the nationality of the regional leader himself suggests that the reason is related to the role of ethnic factors in administration within the republics.